\documentclass[aps,onecolumn,showpacs]{revtex4}

\usepackage{graphicx,color}
\usepackage{amsfonts}
\usepackage{amsmath}
\usepackage{amssymb}
\usepackage[T1]{fontenc}
\usepackage[portuges]{babel}

\newcommand{\bq}{\begin{eqnarray}}
\newcommand{\eq}{\end{eqnarray}}
\newcommand{\be}{\begin{equation}}
\newcommand{\ee}{\end{equation}}
\newcommand{\ei}{\mathrm{e}}
\newcommand{\tr}{\operatorname{tr}}
\newcommand{\ket}[1]{\left| #1 \right\rangle}
\newcommand{\bra}[1]{\left\langle #1 \right|}

\newcommand{\abs}[1]{\left| #1 \right|}

\definecolor{gray}{rgb}{.4,.4,.4}
\definecolor{deepgreen}{rgb}{.1,.6,.3}

\begin{document}

\author{R. Rossi Jr.}
\email{romeu.rossi@ufv.br}
\affiliation{Universidade Federal de Viçosa, Campus Florestal,
C.P 35690-000 - Florestal, MG - Brasil}
\author{J. P. Souza}
\affiliation{Universidade Federal de Viçosa, Campus Florestal,
C.P 35690-000 - Florestal, MG - Brasil}
\author{L. A. M. de Souza}
\affiliation{Universidade Federal de Viçosa, Campus Florestal,
C.P 35690-000 - Florestal, MG - Brasil}
\author{M. C. Nemes}
\affiliation{Departamento de F\'{\i}sica, Instituto de Ciências
Exatas, Universidade Federal de Minas Gerais, C.P. 702, 30161-970,
Belo Horizonte, MG, Brazil}

\title{Multipartite Quantum Eraser}

\begin{abstract}
We study the dynamical entanglement distribution in a multipartite system. The initial state is a maximally entangled two level atom with a single photon field. Next a sequence of atoms are sent, one at the time, and interact with the field. We show that the which way information initially stored only in the field is now distributed among the parties of the global system. We obtain the corresponding complementarity relations in analytical form. We show that this dynamics may lead to a quantum eraser phenomenon provided that measurements of the probe atoms are performed in a basis which maximizes the visibility. The process may be realized in microwave cavities with present technology.
\end{abstract}

\maketitle

\section{Introduction}

In the early days of quantum theory N. Bohr introduced the principle of complementarity. In the original formulation \cite{bohr} the principle states that: "The very nature of quantum theory thus force us to regard the space time co-ordination and the claim of causality, the union of which characterizes the classical theories, as complementary but exclusive features of the description". An alternative statement of complementarity establish the relation between corpuscular and undulatory nature of quantum entities. The so called wave-particle duality can be illustrated in a two-way interferometer, where the apparatus can be set to observe particle behavior, when a single path is taken or wave like behavior, when the interference shows the impossibility to define an actual path.

A modern approach to complementarity includes quantitative relations that are usually written as inequalities relating quantities that represent the possible a priori knowledge of the which path information (predictability) and the ``quality" of the interference fringes (visibility). A considerable number of publications \cite{art1, art2, art3, art4, art5}  contributed to the formulation of the quantitative analysis of the wave-particle duality.

The investigations of systems composed by more than one particle in the context of complementarity are more involved, because there can be entanglement between the particles. Such correlations can give an extra which alternative (path) information about the interferometric possibilities. The quantitative relations for systems composed by two particles were studied in \cite{art6, art7, art8, art9}. Quantitative complementarity relations for multipartite systems have also been formulated. In Ref. \cite{art10} the authors show a quantitative complementarity relation for some specific n-qubit states. In Ref.\cite{art11} a quantitative relation, valid for an arbitrary n-qubit pure state, is given. The relation shows that single particle properties of a component in the n-qubit state depends on the bipartite entanglement that exists between that qubit and the rest of the global system.

The role of entanglement in complementarity is still a mater of investigations. In references \cite{art12, art13, art14} the authors show that the entanglement between an interferometric particle and a probe system can destroy the visibility, because the interaction with a probe system makes the information of which interferometric alternative available. According to the authors, entanglement is the essential key for this phenomena and it is not necessary to call upon Heisenberg's uncertainty principle as it was done in the early discussions between Einstein and Bohr about the recoiling-slit gedanken experiment (a debate on the role of entanglement and uncertainty relations begin \cite{art15, art16, art17, art18, art19}). The authors also pointed out that the which alternative information available on the entangled state can be erased, and consequently visibility recovered, by correlating measurements results of the probe and the interferometric system. Experimental observations of the quantum eraser have been reported in several quantum systems \cite{art20, art21, art22, art23, art24, art25}.

The quantum eraser is not only important for fundamental questions but also for practical applications. As an example quantum eraser was used as a tool for: Channel corrections \cite{art26}, improving cavity spin squeezing \cite{art27},  imaging applications \cite{art28}, experimental entanglement verification \cite{art29}, to quote a few.

In the present work we construct a quantum eraser for a multipartite system with possible experimental implementation. Initially we consider a bipartite system composed by a two level atom, where the atomic energy levels are the interferometric alternatives, and a single cavity mode that works as a probe system. The initial state is maximally entangled, therefore the complete which alternative information is available in the probe system. After, we consider a third subsystem, composed by a set of $n$ two level atoms that interacts, one at the time, with the cavity mode. Such interactions leads to a distribution of the which alternative information on the set of atoms. We used quantitative complementarity relations to show how the information is distributed over the global system. We calculate, for each atom, the measurement result that maximizes the visibility in the interferometric system and show that such atomic measurement sequence also maximizes the loss of which alternative information of the global system. The sequence of measurements work as a quantum eraser of a multipartite system. We also studied the impact the sequence of measurements has on the intrinsic which alternative information of the interferometric system (predictability) and on the information shared with the probe (concurrence). We show that although the global which alternative information is reduced after the calculated measurement sequence, the predictability (of the interferometric system) increases for any number of atoms.

\section{Complementarity Relations and Distinguishability}

In this section we present a definition of the distinguishability and a complementarity relation that will be used in the investigation of the which way information distribution over a multipartite system. The definition of distinguishability, given in Ref.\cite{art9}, for a general state operator of the global system $S+E$, where $S$ represents the system of interest and $E$ other degrees of freedom that interacts with $S$. A general state operator for $S+E$ can be written as

\be\rho_{S,E} = \omega_1 \ket + \bra +
\rho_{E}^{(1)}+ \omega_2 \ket - \bra - \rho_{E}^{(2)} +
\sqrt{\omega_1 \omega_2} (\chi_E \ket - \bra + + h. c.). \ee
where $\rho_{E}^{(1)}$, $\rho_{E}^{(2)}$ and $\chi_E$ are operators in the subsystem $E$.

The distinguishability is given by

\be D = \tr_E \left\{ \abs{\omega_1 \rho_{E}^{(1)} - \omega_2
\rho_{E}^{(2)}} \right\}, \label{defi}\ee.

In physical terms the distinguishability is ``a measure of the possible which path information in an interferometer that Nature can grant us" \cite{art30}.

For a bipartite pure system (composed by a interferometric system and a probe system) in a pure state, the relation between the which path information present on the interferometric system (predictability), on the probe system (concurrence) and the visibility is given by the quantitative complementarity relation \cite{art30}

\be
C^{2}+V^{2}+P^{2}=1.
\ee

For bipartite systems in pure states, the distinguishability can be written as $D=\sqrt{C^{2} + P^{2}}$. This expression shows the single particle and non-classical contributions for the which path information.

\section{The System}

Let us consider the system composed by an atom (we will name it atom zero $S_{a_{0}}$) and an electromagnetic field mode $F$ in the initial state

\be \ket{\psi(0)} = \frac{1}{\sqrt 2}
(\ket{g 1} + \ket{e 0}). \ee

A physical implementation for the state $\ket{\psi(0)}$ can be realized in the system of microwave cavities \cite{art31, art32}.

In the context of quantitative complementarity relations we assume that the atomic energy levels of $S_{0}$ represents two different alternatives in an interferometer and mode $F$ is a probe subsystem that stores the which alternative information of $S_{0}$. The concurrence $C_{S_{0},F}(0)$, visibility $V_{S_{0}}(0)$ and predictability $P_{S_{0}}(0)$ are given by

 \bq
 C^{(0)}_{S_{0},F} = 1,\\
 V^{(0)}_{S_{0}} = 0, \\
 P^{(0)}_{S_{0}} = 0.
 \eq

Notice that as the global system is in a pure state these quantities satisfy the relation $C^{2}+V^{2}+P^{2}=1$ as pointed out in Ref.\cite{art30}. We can also calculate the distinguishability, that in this case can be written as
\be D^{(0)}_{S_{0},F}=\sqrt{C^{2}_{S_{0},F}(0)+P^{2}_{S_{0}}(0)}=1.\ee

In the state $\ket{\psi(0)}$ all information is stored in the probe subsystem $F$ because the global system is in a maximally entangled state. A measurement of $F$ can reveal the which alternative information (which way sorting) or project the subsystem $S_{0}$ in a coherent superposition (quantum erasure sorting) \cite{art9}.

Now let us consider a third subsystem $R$ composed by $N$ two level atoms, prepared in $\ket g$, interacting one at the time with mode $F$. After these sequence of interactions the which alternative information, initially stored in $F$, will be distributed over the $N$ two level atoms and the mode $F$.

The Hamiltonian that governs the interaction of the $k$-th atom is given by
\begin{equation}
\hat{H}^{(k)}= \omega \hat{a}^{\dagger }\hat{a}+
\frac{\omega}{2}\hat{\sigma}_{z}^{(k)}+ g(\hat{a}^{\dagger}\hat{\sigma}_{-}^{(k)}+\hat{a}\hat{\sigma}_{+}^{(k)}), \label{Hamiltonian}
\end{equation}
where $\hat{a}^{\dagger }$ ($\hat{a}$) corresponds to  the creation (annihilation) operator for mode $F$, $\omega $ their frequency, $\hat{\sigma}_{z}^{(k)}=|e^{(k)}\rangle\langle e^{(k)}|-|g^{(k)}\rangle\langle g^{(k)}|$, $\hat{\sigma}_{-}^{(k)}=|g^{(k)}\rangle\langle e^{(k)}|$, $\hat{\sigma}_{+}^{(k)}=|e^{(k)}\rangle\langle g^{(k)}|$ and $g$ the coupling constant for the interaction between $k$-th atom and mode $F$. Here, $\left|g^{(k)} \right\rangle$ and $\left|e^{(k)}\right\rangle$ stand for the ground and the excited states of the $k$-th atom, respectively. Notice that the frequency of atomic transition for all atoms is the same and resonant with the mode frequency.

After the interactions with $n$ atoms the global system is left in the state

\bq \ket {\psi^{(n)}} &=& \frac{1}{\sqrt{2}} (a^n \ket{g} \ket{g_1
g_2 \ldots g_n} \ket{1} + a^{n-1} b \ket{g} \ket{g_1 g_2 \ldots e_n}
\ket{0} + \ldots + a^{i-1} b \ket{g} \ket{g_1 g_2 \ldots g_{i-1}
e_{i} g_{i+1} \ldots g_n} \ket{0}+ \nonumber \\ && + \ldots + a b
\ket g \ket{g_1 e_2 \ldots g_n} \ket{0} + b \ket g \ket{e_1 g_2
\ldots g_n} \ket 0 + \ket e \ket{g_1 \ldots g_n} \ket
0),\label{psin} \eq where $a = \cos\left(\frac{g T}{N}\right)$ and
$b = - i \sin \left(\frac{g T}{N}\right)$, g is the coupling
constant and $T$ is the total time of interaction between mode $F$
and $R$. We consider the same interaction time, with the mode $F$,
for all atoms of $R$. This interaction time is given by $\Delta
t=\frac{T}{N}$ and we also assume that $0<\Delta t<\pi/2g$.To
simplify the notation we can write

\bq \ket {\psi^{(n)}} &=&
\frac{1}{\sqrt 2} (a^n \ket g \ket 0_R \ket 1 + a^{n-1} b \ket g
\ket n_R \ket 0 + \ldots + a b \ket g \ket 2_R \ket 0 + b \ket g
\ket 1_R \ket 0  + \ket e \ket 0_R \ket 0),\eq where $\ket{i_{R}} = \ket{g_1 g_2 \ldots g_{i-1} ~
e_i ~ g_{i+1} \ldots g_n}$ represents a state with an excitations in the i-th atom.

In order to see how the which alternative information is distributed over the global multipartite system, we calculate the distinguishability for the subsystems $S_{0}+F$ and $S_{0}+a_{i}$ where $a_{i}$ represents the subsystem of the $i$-th atom of $R$. In these calculation we follow the definition of distinguishability given in Eq.(\ref{defi}).

To proceed with our calculation we define a normalize state with one excitation in the subsystem $R$

\bq \ket{1_{res}} &=&
\frac{1}{\Gamma} \left( a^{n-1} b \ket{n} + \ldots + a b \ket{2} + b
\ket{1} \right), \eq  with  $\Gamma = 1 - a^{2n}$. We can rewrite the state vector in eq.(\ref{psin}) as

 \bq \ket{\psi^{(n)}} &=&
\frac{1}{\sqrt 2} (a^n \ket g \ket{0_{res}} \ket 1 + \Gamma \ket g
\ket{1_{res}} \ket 0 + \ket e \ket{0_{res}} \ket 0),\eq with
$\ket{0_{res}} = \ket{g_1 g_2 \ldots g_n}$.

The state operator for the global system ($S_{0}+M+R$) is given by

\bq \rho^{(n)} &=& \ket{\psi^{n}}\bra{\psi^{n}} = \frac{1}{2}
\Big( a^{2n} \ket{g ~ 0_{res} ~ 1} \bra{g ~ 0_{res} ~ 1} + \Gamma^2
\ket{g ~ 1_{res} ~ 0} \bra{g ~ 1_{res} ~ 0} + \ket{e ~ 0_{res} ~ 0}
\bra{e ~ 0_{res} ~ 0} + \nonumber \\ && + a^n \Gamma^* \ket{g ~
0_{res} ~ 1} \bra{g ~ 1_{res} ~ 0} + a^n \ket{g ~ 0_{res} ~ 1}
\bra{e ~ 0_{res} ~ 0} + \Gamma \ket{g ~ 1_{res} ~ 0} \bra{e ~
0_{res} ~ 0} + h.c. \Big). \label{rhon} \eq

To calculate the distinguishability we write the reduced state operator of subsystem $S_{0}+F$

\bq
\rho_{S_{0},F}^{(n)} =\tr_R \rho^{(n)} = \frac{1}{2} \Big[ \ket g \bra g \Big( a^{2n}
\ket 1 \bra 1 + \Gamma^2 \ket 0 \bra 0 \Big) + \ket e \bra e \ket 0
\bra 0 + a^n \ket{g 1} \bra{e 0} + h.c. \Big]. \eq

Then we have
\bq D_{S_{0},F}^{(n)} &=& \tr_M \Big\{ \abs{\frac{1}{2} \Big( a^{2n}
\ket 1 \bra 1 + \Gamma^2 \ket 0 \bra 0 \Big) - \frac{1}{2} \ket 0
\bra 0} \Big\} \\ &=&  \frac{1}{2} (a^{2n}
+ \abs{\Gamma^2 - 1}) = a^{2n}.\eq

The distinguishability of $S_{0}+F$ is proportional to the probability to measure the excitation in subsystem $F$ ($P^{(n)}_{F}=a^{2n}/2$). Notice that as the interaction time of each atom is $0<\Delta t<\pi/2g$, the probability $P^{(n)}_{F}$ decreases after each interaction and consequently so does the distinguishability. The which alternative information that was completely stored in $F$ is distributed over the atoms of $R$.

To calculate the distinguishability of $S_{0}+a_{i}$, we write the reduced state operator

\bq \rho_{S_{0},a_i}^{(n)} &=&
\tr_{M,a_1,\ldots,a_{i-1},a_{i+1},\ldots,a_n} \rho^{(n)} = \nonumber
\\ &=& \frac{1}{2} \Big[ \ket g \bra g
\Big( a^{2n} \ket{g_i} \bra{g_i} + (\abs{a^{n-1} b}^2 + \ldots +
\abs{a^{i-2} b}^2 + \abs{a^i b} + \ldots + \abs{b}^2) \ket{g_i}
\bra{g_i} + \abs{a^{i-1} b}^2 \ket{e_i} \bra{e_i} \Big) + \nonumber
\\ && + \ket e \bra e \otimes \ket{g_i} \bra{g_i} + \ket g \bra e
\otimes \frac{1}{\Gamma} \big[ a^{i-1} b \ket{e_i} \bra{g_i} \big]+ h.c.
\Big]. \eq

The distinguishability of $S_{0}+a_{i}$ is given by
 \bq D_{S_{0},a_i}^{(n)} &=& \tr_{a_i} \Big\{ \Big|
\frac{1}{2} a^{2n} \ket{g_i} \bra{g_i} + \abs{a^{i-1} b}^2 \ket{e_i}
\bra{e_i} + (\abs{a^{n-1} b}^2 + \ldots + \abs{a^{i-2} b}^2 +
\abs{a^i b} + \nonumber \\ && + \ldots + \abs{b}^2) \ket{g_i}
\bra{g_i} - \ket{g_i} \bra{g_i} \Big| \Big\} = \abs{a^{i-1} b}^2\label{dist}\eq

The equation
 \be D_{S_{0},F}^{(n)} +
\sum_{i=1}^{n} D_{S_{0},a_i}^{(n)} = a^{2n}+ \abs{a^{n-1} b}^2 + \ldots +
\abs{a^i b} + \ldots + \abs{b}^2 = 1 = D^{(0)}_{S_{0},F},\label{cons}\ee
shows a conservation of distinguishability for the present global system. After the interaction with $n$ atoms of $R$ the which alternative information (that was initially stored in $F$) is distributed over the subsystem $F+R$. The concurrence between subsystems $S_{0}$ and $F+R$ follows the conservation rule for entanglement transfer between qubits reported in Ref.\cite{art33} and is given by $C^{(n)}_{S_{0},F+R}=C^{(0)}_{S_{0},F} = 1$. The equation (\ref{cons}) is also in accordance with the quantitative complementarity relation for multipartite qubit systems given in Ref.\cite{art10}.

The visibility of $S_{0}$ after the $n$ atomic interactions remains with the initial value ($V^{(n)}_{S_{0}} = 0$). To make a detailed analysis, let us write the state operator of eq. (\ref{rhon}) as

\bq \rho^{(n)} &=& \frac{1}{2}
\Big( \rho_{g,g}^{(n)} \ket{g} \bra{g} + \rho_{e,e}^{(n)} \ket{e} \bra{e} + \rho_{e,g}^{(n)} \ket{e} \bra{g} + \rho_{g,e}^{(n)} \ket{g} \bra{e} \Big), \eq

The visibility, defined in Ref.\cite{art9}, is given by:

\bq
V^{(n)}_{S_{0}} = 2 \abs{\tr_{F,R} \rho_{g,e}^{(n)} } = 0,
\eq

where $\rho_{g,e}^{(n)}$ is an operator on the subsystem $F+R$ that multiplies the off diagonal element $\ket{g} \bra{e}$ and can be written as

\bq
\rho_{g,e}^{(n)} = a^n \ket{0_{res} ~ 1} \bra{0_{res} ~ 0} + \Gamma \ket{1_{res} ~ 0} \bra{0_{res} ~ 0}.
\eq

In the calculation of $V^{(n)}_{S_{0}}$ we suppose that after the $n$ atomic interactions no measurement (or non selective measurements) were performed neither on the atoms and neither nor on the mode $F$. Let us consider these assumptions separately. Firstly non selective measurements on the atoms of $R$. The trace on subsystem R deletes the which alternative information stored in $F$ (the assintotic reduced state of $F$ is the vacuum) but it is not a coherent quantum eraser because it also deletes the of diagonal element of the reduced state in $S_{0}$ and consequently its coherence. Secondly, the assumption of non selective measurements on subsystem $F$ is experimentally adequate in microwave cavity contexts where one does not have the direct access to the field, the features of the field mode are obtained through interactions with two level atoms.

Now let us consider (instead of non selective measurements on $R$) that after $n$ atomic interactions a single atom (i-th atom) is measured in the state

\be \ket{M_i} = \alpha_i \ket{g_i} + \beta
\ket{e_i}, \ee
where $\alpha_i = \cos \theta_i$ e $\beta_i = \ei^{i
\phi_i} \sin \theta_i$. In the context of microwave cavities experiments such measurement can be performed using a Ramsey Zone to rotate the atomic state. The vector state $\ket{M_i}$ is an eigenstate of the operator

\be
\hat{\sigma}_{i}=\vec{n}\cdot \vec{\sigma}_{i}
\ee
with $\vec{n}=\left(\sin2\theta_{i} ~ \cos2\theta_{i},~ \sin2\theta_{i}~\sin2\theta_{i},~\cos2\theta_{i}\right)$ and
$\vec{\sigma}=\left(\sigma_{x,i},~\sigma_{y,i},~\sigma_{z,i}\right)$ are the Pauli spin operators of the subsystem of the i-th two level atom.

Formally the quantity $K(\sigma_{i})$, defined in Ref.\cite{art9}, is a quantitative measure of what one can learn about the which alternative information from a measurement of the observable $\sigma_{i}$. For the present system

\be
K(\sigma_{i})= \abs{a^{i-1}b}^{2}\abs{\cos2\theta_{i}}.
\ee

In Ref.\cite{art9} the authors show that the maximum of $K(\sigma_{i})$ is the distinguishability $D$, therefore, the choice of the observable for which $K$ equals $D$ defines a which way sorting. Notice that for $K(\sigma_{i})$ we have a which way sorting when $\theta_{i}=0, ~\pi/2$ (see the distinguishability of the i-th atom in eq.(\ref{dist})). When $\theta_{i}=\pi/4$ a measurement of $\sigma_{i}$ gives no information about the alternative of $S_{0}$.

The visibility after a measurement of the eigenstate $\ket{M_i}$ is given by

\be V^{(n,M_{i})}_{S_{0}}= 2 \abs{\tr_{M,R} P_{i}\rho_{g,e}^{(n)}P_{i} } = \frac{2}{N^{2}_{i}}\abs{\alpha_i \beta^{*}_{i} a^{i-1}b}
\ee

with  $P_i = \mathbb{I}_1 \otimes \ldots \otimes
\mathbb{I}_{i-1} \otimes \ket{M_i}\bra{M_i} \otimes \ldots
\mathbb{I}_n.$ and $N^{2}_{i}=\abs{\alpha_{i}}^{2}\left(2-\abs{a^{i-1}b}^{2}\right)+\abs{\beta_{i}}^{2}\abs{a^{i-1}b}^{2}$.

Notice that a measurement performed in any atom of $R$ increases the visibility of $S_{0}$, except when it represents a which way measurement $\theta_{i}=0, ~\pi/2$.

\section{Multipartite Quantum Eraser}

A quantum eraser measurement aims at removing the which alternative information of a probe system and retrieving the fringe visibility on the system of interest. We show that for the present global system, the which alternative information of $S_{0}$ is distributed over the atoms of $R$ and mode $F$. We have shown above that a single atomic measurement can increase the visibility. In this section we show the quantum eraser measurement for a multipartite system. The multipartite system considered is the global system $S_{0}+F+R$. In the context of microwave cavity experiments, one has no direct access to the cavity mode, but measurements on internal atomic degrees of freedom can be performed. Therefore, in the state of Eq.(\ref{psin}) a projective measurement can be performed on the subsystem $R$, but not on $F$. The quantum multipartite eraser consists in a sequence of measurement results of the internal atomic degrees of freedom (of the atoms in $R$) that maximizes the visibility of $S_{0}$.

To calculate such sequence of measurement results, let us consider projective measurements performed on the state (\ref{psin}). The projetor is given by

\be P = P_1 \otimes \ldots
\otimes P_n, \ee
where

\be P_i = \mathbb{I}_1 \otimes \ldots \otimes
\mathbb{I}_{i-1} \otimes \ket{M_i}\bra{M_i} \otimes \ldots
\mathbb{I}_n. \ee

Notice that $P$ acts only on the subsystem $R$ and represents a projective measurement of each atomic internal state for all the atoms in $R$. After the $n$ projective measurements the normalized global state vector is given by

\be \ket
\psi^{(n,M)} = \frac{1}{N}(A \ket g \ket M \ket 1 + B \ket g \ket M \ket 0
+ C \ket e \ket M \ket 0), \ee where $\ket M = \ket{M_1} \ldots \ket{M_n}$ and  \bq N &=& \sqrt{\abs{A}^2 +
\abs{B}^2 + \abs{C}^2} \nonumber \\ A &=& \frac{1}{\sqrt 2}
\left(a^n \prod_{i=1}^{n} \alpha_i \right) \nonumber \\ B &=&
\frac{1}{\sqrt 2} \left(b \sum_{i = 1}^{n} \left[ a^{i-1}
\frac{\beta_i}{\alpha_i}
\left( \prod_{j=1}^{n} \alpha_j \right) \right] \right)\nonumber \\
C &=& \frac{1}{\sqrt 2} \left( \prod_{i=1}^{n} \alpha_i \right).\eq

Notice that after the measurements the atoms of $R$ are not entangled with the subsystem $S_{0}+F$, but there is entanglement between $S_{0}$ and $F$. The visibility, predictability and concurrence after the measurements are given by:

\bq V^{(n,M)}_{S_{0}} &=& \frac{2 \abs{B C^*}}{N^2}
\\ P^{(n,M)}_{S_{0}} &=& \frac{\abs{C}^2
- \abs{B}^2-\abs{A}^2}{N^2} \\ C^{(n,M)}_{S_{0},F} &=& \frac{2 \abs{A C}}{N^2}, \eq

The reduced state on the bipartite system $S_{0}+F$ is a pure state, therefore we can write the complementarity relation
\be
\left(C^{(n,M)}_{S_{0},F}\right)^2 +\left(P^{(n,M)}_{S_{0}}\right)^2 +\left(V^{(n,M)}_{S_{0}}\right)^2 = 1, \label{complem1}\ee

\begin{figure}[h]
\centering
  \includegraphics[scale=1]{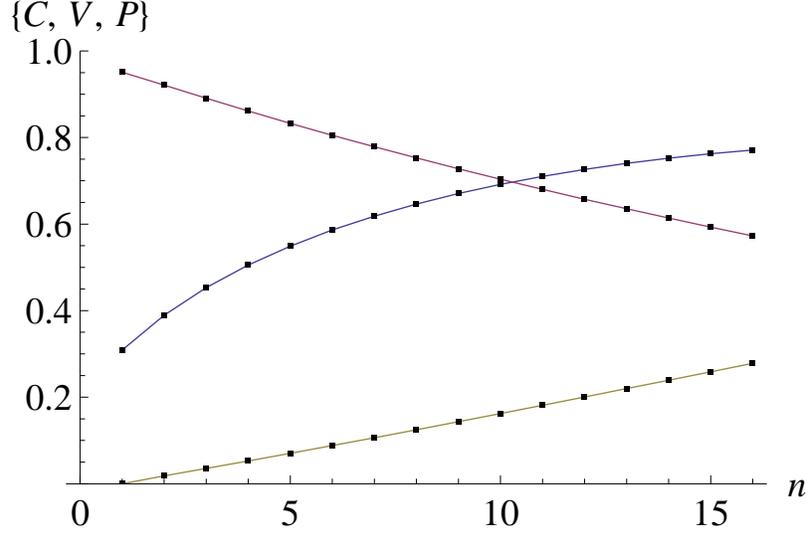}\\
\caption{Concurrence (purple), visibility (blue) and predictability
(beige) as a function of \emph{n}. Parameters: $g T = 2 \pi$, $N =
20$ atoms and the coefficients $\alpha_i$ and $\beta_i$ are given by
the maximization procedure.}
\end{figure}

The $C^{(n,M)}_{S_{0},F}$, $P^{(n,M)}_{S_{0}}$ and $V^{(n,M)}_{S_{0}}$ depend on the coefficients $\alpha_i$ and $\beta_i$ that are determined by the results of the atomic measurements. In Fig.1 we show curves of $C^{(n,M)}_{S_{0},F}$, $P^{(n,M)}_{S_{0}}$ and $V^{(n,M)}_{S_{0}}$ for results of atomic measurements that maximize the visibility. The maximization was obtained by a numerical process. The information initially stored in $F$ was distributed after the interactions with $R$ ($R$ in Fig.1 is composed by $N=20$ atoms). Therefore each atom of $R$ stores some which alternative information. The information in each atom is maximally erased after the measurement of the eigenstate $\ket{M}_{i}$ (of the observable $\sigma_{i}$), that is determined by the visibility maximization procedure. The visibility increases with the number of measured atoms ($n$) because after each measurement more which alternative information is erased. The predictability also increases with the number of measured atoms in Fig.1, this indicates that the projective atomic measurements increase the which way information in subsistem $S_{0}$. The concurrence decreases with $n$, this is due to the conservation rule for entanglement transfer between qubits reported in Ref.\cite{art33}. Interaction with $n$ atoms reduces the initial entanglement on the subsystem $S_{0}+F$, it is distributed over the global multipartite system.

The variation of the global which way information after the measurements is an essential ingredient for the increase of the visibility. To clarify the relation between this two quantities let us consider the quantitative complementarity relation before and after the measurements.

Before the measurements, the global system is a multipartite system composed by $N+2$ qubits in a generalized W state in Eq.(\ref{psin}). In Ref.\cite{art10} the authors show a complementarity relation for a 3-qubit system in the generalized W state

\be \sum_{k\neq j}C^{2}_{k,j}+V^{2}_{k}+P^{2}_{k} = 1,\ee
where $C_{k,j}$ is the concurrence between the $k$-th and the $j$-th qubits, $V_{k}$ and $P_{k}$ are respectively the visibility and the predictability of the $k$-th qubit. In this complementarity relation the terms $C_{k,j}$ and $P_{k}$ are related to the which way information.

For the global system $S_{0}+F+R$ before the measurement we can write

\be
\left(C_{S_{0},F}\right)^2 +\sum_{i=1}^{n}\left(C_{S_{0},a_i}\right)^2 + \left(P_{S_{0}}\right)^2 +\left(V_{S_{0}}\right)^2 = 1 \label{complem2}, \ee

From equations (\ref{complem1}) and (\ref{complem2}) we can write

\be \left(C_{S_{0},F}\right)^2 +\sum_{i=1}^{n}\left(C_{S_{0},a_i}\right)^2 + \left(P_{S_{0}}\right)^2 - \left(C^{(n,M)}_{S_{0},F}\right)^2 - \left(P^{(n,M)}_{S_{0}}\right)^2 = \left(V^{(n,M)}_{S_{0}}\right)^2 - \left(V_{S_{0}}\right)^2.\label{complem3}\ee

The left-hand side of equation (\ref{complem3}) is related to the variation of which way information, and the right-hand side is related to the variation of the visibility. Therefore it is clear that if the measurements increase the visibility ($V^{(n,M)}_{S_{0}} > V_{S_{0}}$) part of the global which way information is lost.

In the vector state of equation (\ref{psin}) the predictability of $S_{0}$ is $P^{(n)}_{S_{0}}=0$ and the subsistem $F+R$ has the complete which way information as shown in equation (\ref{cons}). For convenience we define $D^{(n)}_{T} $ as the total distinguishability

\be
D^{(n)}_{T} = \sum^{N}_{i\neq S_{0}}C^{2}_{S_{0},i} = a^{2n}+ \abs{a^{n-1} b}^2 + \ldots +
\abs{a^i b}^{2} + \ldots + \abs{b}^2 =D_{F}^{(n)} +
\sum_{i=1}^{n} D_{a_i}^{(n)} = 1. \ee

After the measurement the distinguishability of the subsystem and can be written as

\be
D^{(n,\textbf{M})}_{T}= \sqrt{\left(C^{(n,\textbf{M})}_{S_{0},F}\right)^{2}+\left(P^{(n,\textbf{M})}_{S_{0}}\right)^{2}}.
\ee

The variation of the total distinguishability before and after the measurements can be written as

\be
\Delta D_{T} = D^{(n,\textbf{M})}_{T} - D^{(n)}_{T} =\sqrt{\left(C^{(n,\textbf{M})}_{S_{0},F}\right)^{2}+\left(P^{(n,\textbf{M})}_{S_{0}}\right)^{2}} - 1= \sqrt{1-\left(V^{(n,\textbf{M})}_{S_{0}}\right)^{2}} - 1,
\ee
$\Delta D_{T} $ is the variation of the global distinguishability before and after the measurements.

Notice that, as $D^{(n)}_{T}=1$ the complementarity relations show us that any sequence of measurements that increases the visibility reduces the global which way information. Therefore, the sequence of measurements that maximizes the visibility erases as much which way information as possible from the global system.

\begin{figure}[h]
\centering
  \includegraphics[scale=1]{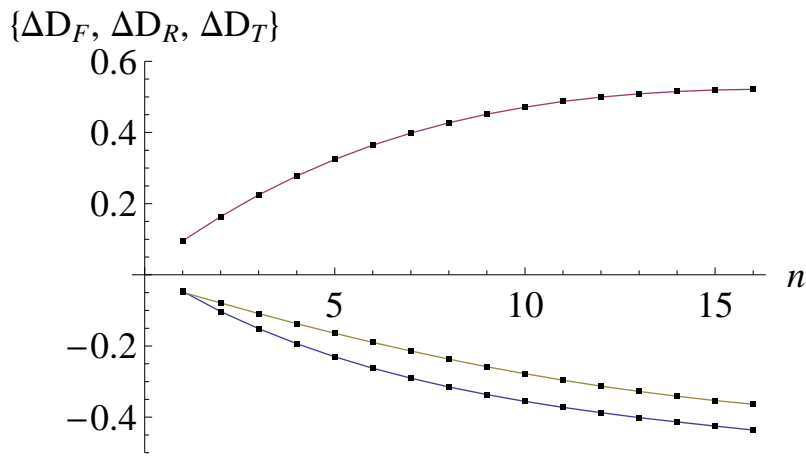}\\
\caption{$\Delta D_F$ (purple), $\Delta D_R$ (blue) and $\Delta D_T$
(beige) as a function of \emph{n}. Parameters: $g T = 2 \pi$, $N =
20$ atoms and the coefficients $\alpha_i$ and $\beta_i$ are given by
the maximization procedure.}
\end{figure}

In Fig.2 it is shown $\Delta D_{T}$ after a sequence of measurements, of $n$ atoms of $R$, that maximizes the visibility of $S_{0}$. It is also shown the variation of the information of subsystem $S_{0} + F$, that is given by $\Delta D_{F}=\sqrt{1-\left(V^{(n,\textbf{M})}_{S_{0}}\right)^{2}}-a^{2n}$. Notice that the variation $\Delta D_{F}$ is positive. This indicates that although the global information is lost after the measurement process, the information retained in the subsystem $S_{0} + F$ increases when compared with its value before the atomic measurements.

In conclusion, we study the entanglement distribution in a multipartite system that is composed by an electromagnetic field mode and $N+1$ two level atoms. The entanglement initially in the bipartite system $S_{0}+F$ is distributed over the multipartite system by interactions between the $N$ atoms and the field mode. In the context of complementarity relations we consider the subsystem $S_{0}$ as the interferometric system and $F$ as the probe system. After the interactions between $F$ and the $N$ atoms, the which way information initially stored in the probe, is distributed over the multipartite system. We show that the distributed which way information can be maximally erased by measurements performed on the $N$ interacting atoms in a basis which maximizes the visibility. The present multipartite quantum eraser can be realized in microwave cavity system \cite{art31,art32}.

\end{document}